\documentstyle[12pt]{article}
\begin{document}
\newcommand{\be}{\begin{equation}}
\newcommand{\ee}{\end{equation}}
\newcommand{\bea}{\begin{eqnarray}}
\newcommand{\eea}{\end{eqnarray}}
\newcommand{\beas}{\begin{eqnarray*}}
\newcommand{\eeas}{\end{eqnarray*}}
\newcommand{\ba}{\begin{array}}
\newcommand{\ea}{\end{array}}
\newcommand{\nn}{\nonumber}
\newcommand{\nonu}{\nonumber}
\newcommand{\bfg}{\begin{figure}}
\newcommand{\efg}{\end{figure}}
\newcommand{\tl}{\tilde}
\newcommand{\rar}{\rightarrow}
\newcommand{\lar}{\leftarrow}
\newcommand{\lrar}{\longrightarrow}
\newcommand{\llar}{\longleftarrow}
\newcommand{\fr}{\frac}
\newcommand{\pa}{\partial}
\newcommand{\mb}{\mbox}
\newcommand{\lft}{\lefteqn}
\newtheorem{th}{Theorem}
\newtheorem{lm}{Lemma}
\newtheorem{cl}{Corollary}
\newtheorem{df}{Definition}
\newtheorem{E}{Example}
\newcommand{\bth}{\begin{th}}
\newcommand{\eth}{\end{th}}
\newcommand{\blm}{\begin{lm}}
\newcommand{\elm}{\end{lm}}
\newcommand{\bcl}{\begin{cl}}
\newcommand{\ecl}{\end{cl}}
\newcommand{\bdf}{\begin{df}}
\newcommand{\edf}{\end{df}}
\newcommand{\brk}{\begin{rm}}
\newcommand{\erk}{\end{rm}}
\newcommand{\hs}{\hspace}
\newcommand{\vs}{\vspace}
\newcommand{\hst}{\hspace*}
\newcommand{\vst}{\vspace*}
\newcommand{\lb}{\label}
\newcommand{\nl}{\newline}
\newcommand{\np}{\newpage}
\newcommand{\om}{\omega}
\newcommand{\Om}{\Omega}
\newcommand{\al}{\alpha}
\newcommand{\bt}{\beta}
\newcommand{\dt}{\delta}
\newcommand{\eps}{\epsilon}
\newcommand{\veps}{\varepsilon}
\newcommand{\ld}{\lambda}
\newcommand{\Ld}{\Lambda}
\newcommand{\gm}{\gamma}
\newcommand{\Gm}{\Gamma}
\newcommand{\sg}{\sigma}
\newcommand{\Sg}{\Sigma}
\newcommand{\bib}{\bibitem}
\newcommand{\ct}{\cite}
\newcommand{\rf}{\ref}
\newcommand{\abschnitt}[1]{\par \noindent {\large {\bf {#1}}} \par}
\newcommand{\subabschnitt}[1]{\par \noindent
                                          {\normalsize {\it {#1}}} \par}
%
\newcommand{\vX}{{\bf X}}
\newcommand{\vnab}{\mbox{\boldmath $\nabla$}}
\newcommand{\nab}{\nabla}
\newcommand{\rcn}{\vr\cdot\vnab}
\newcommand{\rcB}{\vr\cdot\vB}
\newcommand{\rcE}{\vr\cdot\vE}
\newcommand{\LcE}{\vL\cdot\vE}
\newcommand{\LcB}{\vL\cdot\vB}
\newcommand{\bigtr}{\bigtriangleup}
\newcommand{\vr}{{\bf r}}
\newcommand{\vE}{{\bf E}}
\newcommand{\vH}{{\bf H}}
\newcommand{\vP}{{\bf P}}
\newcommand{\vM}{{\bf M}}
\newcommand{\vA}{{\bf A}}
\newcommand{\vD}{{\bf D}}
\newcommand{\vB}{{\bf B}}
\newcommand{\vS}{{\bf S}}
\newcommand{\vL}{{\bf L}}
\newcommand{\vk}{{\bf k}}
\newcommand{\vn}{{\bf n}}
\newcommand{\vK}{{\bf K}}
\newcommand{\vV}{{\bf V}}
\newcommand{\CMP}[1]{{\em Comm. Math. Phys.}\ {\bf #1}}
\newcommand{\CQG}[1]{{\em Class. Quantum Grav.}\ {\bf #1}}
\newcommand{\EL}[1]{{\em Europhys. Lett.}\ {\bf #1}}
\newcommand{\IJMPB}[1]{{\em Int. J. Mod. Phys.} B\ {\bf #1}}
\newcommand{\JMP}[1]{{\em J. Math. Phys.}\ {\bf #1}}
\newcommand{\JOP}[1]{{\em J. of Operator Theory}\ {\bf #1}}
\newcommand{\JPL}[1]{{\em J. Phys.} (Paris)\ {\bf #1}}
\newcommand{\MPLA}[1]{{\em Mod. Phys. Lett.} A\ {\bf #1}}
\newcommand{\MPLB}[1]{{\em Mod. Phys. Lett.} B\ {\bf #1}}
\newcommand{\NC}[1]{{\em Nuovo Cimento}\ {\bf #1}}
\newcommand{\NP}[1]{{\em Nucl. Phys.}\ {\bf #1}}
\newcommand{\PH}[1]{{\em Physica}\ {\bf #1}}
\newcommand{\PL}[1]{{\em Phys. Lett.}\ {\bf #1}}
\newcommand{\PR}[1]{{\em Phys. Rev.}\ {\bf #1}}
\newcommand{\PRB}[1]{{\em Phys. Rev.} B\ {\bf #1}}
\newcommand{\PRE}[1]{{\em Phys. Rep.}\ {\bf #1}}
\newcommand{\PRL}[1]{{\em Phys. Rev. Lett.}\ {\bf #1}}
\newcommand{\RMP}[1]{{\em Rev. Mod. Phys.}\ {\bf #1}}
\hyphenation{me-di-um  as-su-ming pri-mi-ti-ve pe-ri-o-di-ci-ty}
\hyphenation{in-te-gral e-le-c-tro-ma-g-ne-tic e-qua-ti-ons}
\title{
\hfill{\small PRA-HEP 93/10}\\
\hfill{\small June 1993}\vst{1cm}\\
\sc
Inward and outward integral equations and the KKR method for
photons
\vst{0.8cm}}
\author{\sc Alexander Moroz\thanks{e-mail address :
{\tt moroz@fzu.cz}}\
\thanks{Address after September 1, 1993:
Div.  of Theor. Physics, IPN,
Univ. Paris-Sud, F-91 406 Orsay Cedex, France.}\\
\protect\normalsize
\it Institute of Physics CAS,
\it Na Slovance 2\\
\protect\normalsize\it CZ-180 40 Prague 8,
\it Czech Republic}
\date{}
\maketitle
\begin{center}
{\large\sc abstract}
\end{center}
\baselineskip 14pt
{
\footnotesize
\noindent
In the case of electromagnetic waves it is necessary to distinguish between
inward and outward on-shell integral equations. Both kinds of
equation are derived. A correct implementation of the photonic KKR method
then requires the inward equations and it follows directly from them.
A derivation of the KKR method from
a variational principle is also outlined. Rather surprisingly, the variational
KKR method cannot be entirely written in terms of surface integrals unless
permeabilities are piecewise constant.
Both kinds of photonic KKR method use the standard structure constants
of the electronic KKR method and hence allow for a direct numerical
application. As a by-product, matching rules are obtained
for derivatives of fields on
different sides of the discontinuity of permeabilities.}

\vspace*{2cm}

\begin{center}
{\em (To appear in J. Phys. : Cond. Matter)}
\end{center}
\thispagestyle{empty}

\baselineskip 20pt
\newpage
\noindent
\section{Introduction}
This paper is first from an intended series of papers devoted to
problems of photonic band gap and the propagation of electromagnetic waves
in periodic dielectric structures. The question of the existence of a
photonic band gap can be rephrased by asking whether it is possible
to prepare a {\em perfect mirror from a transparent material} by
drilling (in an appropriate way) holes to it.
The existence of a photonic band gap has been suggested theoretically \ct{Jo}
and recently also been observed experimentally \ct{YG}.
Nevertheless, there has been no
calculation of band structure by more reliable methods,
such as the KKR (Kohn-Korringa-Rostocker)  or APW (augmented plane wave)
methods. Even worse, a derivation of the photonic KKR method along lines
of \ct{KKR} has not yet been given . Also, both theoretically and
numerically, only a limited range of dielectric lattices has
so far been considered  - diamond like lattice, fcc and simple cubic
lattices \ct{HCS}.
In all papers  referred to, ``{\em dielectric atoms}"
(a compact piece of dielectric in which (and only in which) permeabilities are
allowed to differ from their host or average values)
have been assumed to be spherically symmetric.

In attempting a calculation of the band structure of photons on various
dielectric lattices with spherical atoms
we have  tried to generalize the scalar bulk KKR method
\ct{KKR} to the form appropriate for photons \ct{AK}.
In the course of its derivation we have found that there  persists some
confusion in the photonic KKR method. This originates in the fact
that, unlike the scalar case, in the case of photons fields
can change {\em discontinuously} over {\em discontinuities} of
electric and magnetic permeabilities.
This is the essential difference between the Schr\"{o}dinger and
the Maxwell equations. Consequently,
derivatives of fields may be singular and inward and outward limits of
volume integrals over atoms may differ.
Therefore, one has to carefully distinguish between ``{\em inward}" and
``{\em outward}"
formulations, i.\,e., whether, in a given on-shell \ct{ON} surface integral,
inward or outward limits of fields and their derivatives at the
atom boundary are taken.
In particular, we shall show that the photonic KKR method proposed
in \ct{LWA} does not serve the purpose.
In \ct{LWA}, fields in  integral equations are {\em outward} limits of fields
with respect to the atom boundary, while the scalar product
with {\em inward} limits of fields is taken.

These subtleties in derivation of the photonic KKR
method have stimulated us to write this paper. We hope
that this derivation may be both instructive and interesting by itself.
A more complete treatment of this subject, together with numerical
results on photonic band structure, will be given elsewhere \ct{AK}.

\section{Integral equations for electromagnetic potentials}
\lb{intreq}
\subsection{Preliminaries}
\lb{prelim}
We shall consider the {\em stationary macroscopic} Maxwell equations
in a {\em non-conducting} medium with the current and
charge densities equal to zero \ct{BW}.
We shall only consider the simplest isotropic case, $\vD(\vr) =
\eps(\vr)\vE(\vr)$ and $\vB(\vr)=\mu(\vr)\vH(\vr)$ where
$\vD(\vr)$ ($\vB(\vr)$) is the electric (magnetic) induction.
We shall confine ourselves to monochromatic waves
to avoid non-local time relation between $\vD(\vr)$ and $\vE(\vr)$
or $\vB(\vr)$ and $\vH(\vr)$. We shall also allow for {\em complex}
 permeabilities, i.\,e., for an absorption.
In such medium the Maxwell equations are  symmetric under
\be
\vE(\vr)\rar\vH(\vr),\hs{0.5cm}\vH(\vr)\rar -\vE(\vr),\hs{0.5cm}
\eps(\vr)\rar\mu(\vr).
\lb{symm}
\ee
They can be written as
\be
\vnab\times{\bf H}(\vr) + i\,\fr{\om}{c}\,\eps(\vr)\,{\bf E}(\vr) = 0,
\lb{ME1}
\ee
\be
\vnab\times{\bf E}(\vr)- i\,\fr{\om}{c}\,\mu(\vr)\,{\bf H}(\vr) = 0,
\lb{ME2}
\ee
\be
\vnab\cdot\vD(\vr)= \vnab\cdot\vB(\vr) = 0.
\lb{con}
\ee
Now, by combining (\rf{ME1}) and (\rf{ME2}) one gets
\bea
\lft{\vnab\times[\vnab\times \vE(\vr)] -(\om/c)^2\eps(\vr)\mu(\vr)
\vE(\vr)}\hs{2cm}\nn\\
&&-\mu^{-1}(\vr)[\vnab\mu(\vr)]\times[\vnab\times\vE(\vr)]=0,
\lb{ME*}
\eea
as well as
\bea
\lft{\vnab\times[\vnab\times \vH(\vr)] -(\om/c)^2\eps(\vr)\mu(\vr)\vH(\vr)}
\hs{2cm}\nn\\
&&-\eps^{-1}(\vr)[\vnab\eps(\vr)]\times[\vnab\times\vH(\vr)]=0.
\lb{ME6*}
\eea
Due to the symmetry (\rf{symm}) of the Maxwell equations,
frequently only a single equation will be written  from a conjugate pair.
If $\mu(\vr)$ is set to be a constant, $\mu(\vr)=\mu_o$,
and one parametrizes $\mu_o\eps(\vr)$ into its constant uniform value
$\mu_o\eps_o$ and a spatially varying part $v(\vr)=\mu_o[\eps(\vr)-\eps_o]$,
one finds a striking resemblence of (\rf{ME*}) and the Schr\"{o}dinger
equation, with $v(\vr)$ playing the role of a potential. Theoretically, this
analogy serves as a main motivation for the search of a photonic band gap.

We shall see that to solve the Maxwell equations, even in the case when the
potentials $\eps(\vr)$ and $\mu(\vr)$ can
be written as a sum of functions of a single coordinate,
 is a {\em non-trivial}  task
because they {\em do not separate}. The reason is that any solution
of, say, (\rf{ME*}) has to satisfy the initial conditions (\rf{con})
for the Maxwell equations in a dielectric. Provided that $\eps(\vr)$
is a non-trivial function of all coordinates then (\rf{con})
mixes all components of $\vE(\vr)$ together,
\be
\vnab\cdot\vE(\vr) =-\fr{1}{\eps(\vr)}\,[\vnab\eps(\vr)\cdot\vE(\vr)].
\lb{inc}
\ee
Similarly for $\vH(\vr)$ in the case of non-trivial $\mu(\vr)$.

We shall now turn on to the derivation of integral equations for
electromagnetic fields.
We shall consider either finite or infinite region $\Om$
with given boundary conditions imposed and with one
dielectric atom $V_s$ inside it. The generalization to several atoms is
trivial. Since the group of lattice translations is Abelian,
the Bloch theorem extends straightforwardly to vector
wave functions. Hence, by setting $\Om$ to be a primitive cell
and with generalized periodic boundary conditions imposed, one can
directly turn on to the study of electromagnetic wave propagation in
dielectric lattices.

\subsection{``Outward'' integral equations}
The first attempt to derive integral equations for
electromagnetic fields in a dielectric in connection with the photonic
KKR method is discussed in \ct{LWA}. The derivation
is basically due to Morse \ct{KKR} and starts as follows.
Provided $\vr$ is restricted to the {\em interstitial} region,
$\vr\in\Om\backslash V_s$,  the electric
(magnetic) field $\vE(\vr)$ ($\vB(\vr)$) satisfies the equation
\be
(\vnab^2 +\sg^2 )\vE(\vr) = 0,
\lb{free}
\ee
where $\sg^2 =\om^2\eps_o\mu_o$.
Let $G_{\sg}(\vr,\vr')$ be the Green function of the
Helmholtz equation in  $\Om$ with suitable boundary
conditions imposed, defined by
\be
(\vnab^2 +\sg^2)G_\sg (\vr ,\vr ') = \delta (\vr -\vr ').
\lb{green}
\ee
By using the defining equation (\rf{green}) for  $G_\sg (\vr, \vr ')$
one is tempting to write  the following on-shell integral equation in the
interstitial region,
\bea
\lft{\vE (\vr) = \int_{\Om\backslash V_s} \delta(\vr -\vr ')\vE(\vr ')
\,d^3\vr '=}\nn\\
&& =\int_{\Om\backslash V_s} [
\vE(\vr ') \vnab '^{2} G_\sg (\vr, \vr ') -
G_\sg (\vr,\vr ')\vnab '^{2}\vE (\vr ')]\,d^3\vr ',
\lb{zero}
\eea
which can be obviously rewritten as the surface integral,
\be
\vE (\vr)= \oint_{\pa(\Om\backslash V_s)} [\vE(\vr ')\,
(d\vS '\cdot\vnab ') G_\sg (\vr,\vr ') -
G_\sg(\vr, \vr ')\,(d\vS '\cdot\vnab ')\vE (\vr ')].
\lb{other}
\ee
Now, assuming $\Om $ to be a primitive cell, then
due to the periodicity of $G_\sg (\vr,\vr ')\,\vE(\vr ')$,
 the integral over
the {\em cell} boundary vanishes and one is left with the surface integral
over $\pa V_s$.  The same equation also holds for $\vB(\vr)$.
Thus, formally, both $\vB(\vr)$ and $\vE(\vr)$ satisfy  the same
integral equation (\rf{other}) outside a dielectric atom $V_s$.
What makes the difference between them are the
matching conditions across a boundary of the atom.

However, there is one subtle point here. In (\rf{other}) an
{\em outward} limit of the electric field $\vE(\vr)$ and its derivatives
is taken.
On the other hand, the KKR method requires knowledge  of either interior
or exterior solutions. However, the exterior problem is much more
involved and {\em cannot} be  solved usually. Only {\em interior} solution
can (sometimes) be
obtained in an explicite form, supposing that there are some
symmetries of the problem (such as spherical symmetry, for example).
Therefore, despite that in
the surface  integral (\rf{other}) the {\em outward} limit of the
electric field $\vE(\vr)$ and its derivatives is taken,
usually only the {\em inward} limit of field and its derivatives is at
our disposal. Now, our main task is to find integral equations for
electromagnetic fields in $\Om$ in terms of the {\em inward limit} of fields
and their derivatives at the boundary of $V_s$.
Provided that $\eps(\vr)$ is
{\em discontinuous}, the discontinuity of the normal component $\vE(\vr)$ at
the atom boundary is
\be
\vn\cdot[\vE_+(\vr)-\vE_-(\vr)]=\fr{v(\vr)}{\eps_o}\,\vE_-(\vr)
\lb{disc}
\ee
where $\vn$ is the unit normal vector at the discontinuity.
Thus, equations (\rf{ME*}-\rf{ME6*}) inevitably contain {\em singular} terms.
This makes a relation between inward and outward equations non-trivial.
It is this point that was not taken into account in \ct{LWA}.

It what follows it will be assumed that on discontinuities
of $\eps(\vr)$ the side limits of of $\eps(\vr)$ and its derivatives
are well defined. The same will be assumed for limiting values of
fields and their derivatives.

\subsection{``Inward" integral equations}
In trying to find the ``{\em inward}" integral equations one can try
to establish a direct relation between fields and their
derivatives on both sides of the atom boundary and rewrite (\rf{other})
in terms of inward limits. It can be shown that like-side limits
of $\eps(\vr)$ determine relation of the side limits of $\vE(\vr)$
(\rf{disc}), the side limits of {\em derivatives} of $\eps(\vr)$ do so for
the side limits of derivatives of $\vE(\vr)$ (see Appendix).
Before proceeding this way,  the ``inward" integral equations will be
derived by introducing electromagnetic potentials. This way seems to be
easier and safer in treating  singularities that  appear.

If the additive material relations are used
\be
\vD(\vr)=\eps_o\vE(\vr)+4\pi\vP(\vr),\hs{0.8cm}
\vB(\vr)=\vH(\vr)+4\pi\vM(\vr),
\ee
the Maxwell equations take on the following form
\be
\vnab\times{\bf B} + i\om\eps_o\,{\bf E} = 4\pi\tilde{\bf j},
\lb{bum1}
\ee
\be
\vnab\times {\bf E} -i\om\,{\bf B} = 0,
\lb{bum2}
\ee
\be
\vnab\cdot{\bf E} = 4\pi\tilde{\rho},
\lb{ME3}
\ee
\be
\vnab\cdot{\bf B} = 0.
\lb{ME4}
\ee
The current and charge density are defined by the relations
\be
\tilde{\bf j} = -i\om\,{\bf P}+c\vnab\times\vM(\vr),
\ee
\be
\tilde{\rho} = -\fr{1}{\eps_o}\,\vnab\cdot{\bf P},
\ee
where $\bf P(\vr)$ ($\vM(\vr)$) is the {\em electric (magnetic) polarization},
\be
{\bf P}(\vr) = \fr{v(\vr)}{4\pi}\,\vE(\vr),\hs{0.8cm}
\vM(\vr) =\fr{\mu(\vr)-1}{4\pi}\vH(\vr).
\lb{conv}
\ee
The equations (\rf{bum1}-\rf{ME4}) are formally identical to the equations
for the electromagnetic field in a vacuum. By introducing a vector
potential $\bf A(\vr)$ and a scalar potential $\phi(\vr)$ such that
\be
{\bf B}(\vr) =\vnab\times{\bf A}(\vr),
\lb{defb}
\ee
\be
{\bf E}(\vr) = i\om{\bf A}(\vr)-\vnab\phi(\vr),
\lb{defe}
\ee
with the Lorentz gauge condition imposed,
\be
\vnab\cdot{\bf A}(\vr)=i\om\,\eps_o\,\phi(\vr),
\lb{LC}
\ee
equations for $\bf A(\vr)$ and $\phi(\vr)$ can be {\em decoupled}
\ct{BW}.

In order to simplify our discussion  the dielectric medium
 will be assumed to be {\em magnetically isotropic} and from now on
we shall set  $\mu(\vr) =1$ as well as $c=1$. For the potentials one gets
\be
\vnab^2{\bf A}(\vr) +\sg^2{\bf A}(\vr)=i\om\, v(\vr)\,{\bf E}(\vr),
\lb{apot}
\ee
\be
\vnab^2\phi(\vr) +\sg^2\phi(\vr) =-\vnab\cdot{\bf E}(\vr).
\lb{phipot}
\ee
The integral equations for $\bf A(\vr)$ and $\phi(\vr)$ are
\be
{\bf A}(\vr) = i\, \om \int_{V_s} G_{\sg}
(\vr,\vr ')\, v(\vr ')\, \vE (\vr ') \,d^3\vr ',
\lb{inteq}
\ee
\be
\phi (\vr) = -\int_{V_s} G_{\sg}
(\vr,\vr ')\,[\vnab '\cdot\vE (\vr ')]\,d^3\vr '.
\lb{iphipot}
\ee
As we have mentioned above, due to the presence of a singular shell at the
boundary of $V_s$ the volume integrals over $V_s$ have to be defined
carefully as an {\em outward limit} through measurable sets $\Sg_n$,
$\Om\supset\Sg_n\supset V_s$, $\lim_{n\rar\infty} \Sg_n \searrow V_s$.
The Gauss theorem requires the integrand to be at least
a continuous function. Provided that a discontinuity appears in some region
$\Sg$ to which it is going to be applied one  proceeds
as follows. The region $\Sg$ is split (if possible) into non-overlapping
subregions $\Sg_j$, $\Sg =\cup_j \Sg_j$, such that in any $\Sg_j$
the hypotheses of the Gauss theorem are satisfied. Then, the Gauss theorem
is applied to each $\Sg_j$ separately. As mentioned above, inward and
outward limits of volume integrals with a singular integrand at the boundary
will differ in general. First example provides (\rf{iphipot}).
By using
\be
\vnab\cdot [v(\vr)\vE (\vr)]= -\eps_o \vnab\cdot\vE(\vr)
\lb{id}
\ee
one finds
\bea
\lft{\phi(\vr) := -\int_{V_{s+}} G_{\sg}
(\vr,\vr ')\, [\vnab '\cdot \vE (\vr ')] \,d^3\vr '=}\nn\\
&&- \int_{V_{s-}} G_{\sg}
(\vr,\vr ')\, [\vnab '\cdot\vE (\vr ')] \,d^3\vr '
- \fr{1}{\eps_o}\oint_{\pa V_{s-}} G_{\sg}(\vr,\vr ')\,v(\vr ')
(\vE(\vr')\cdot d\vS '),\hs{0.7cm}
\lb{unsym}
\eea
where $V_{s+}$ ($V_{s-}$) means that the outward (inward) limit is taken.
Here, the surface integral exactly correponds to a delta function
singularity which appears due to the discontinuity (\rf{disc}) of the normal
component of $\vE(\vr)$.  Now, it is tranparent that, provided
the inward limit in (\rf{iphipot}) is taken, the
resulting integral expressions (\rf{inteq}-\rf{iphipot}) for electromagnetic
potentials do not satisfy the Lorentz condition (\rf{LC}).

The off-shell integral equation for $\vE(\vr)$ can be directly written
by using the defining relations (\rf{defe}) and (\rf{inteq},\rf{iphipot}),
\bea
\lft{\vE(\vr)= -\om^2\,
\int_{V_s} G_{\sg}(\vr ,\vr ')\,v(\vr')\,\vE (\vr ')\,d^3\vr '-}
\hs{2.5cm}\nonumber\\
&&\mbox{} -\int_{V_{s+}} [\vnab'
G_{\sg}(\vr ,\vr ')] [\vnab'\cdot\vE (\vr ')] \,d^3\vr '.
\lb{entq}
\eea
To find an on-shell integral equation for $\vE(\vr)$
we shall make repeated use of (\rf{ME*}) together with the identity (\rf{id})
and with (\rf{green}) satisfied by $G_\sg(\vr,\vr')$.
For a given region $\Sg$,  $\Sg\subset V_s$, one finds in the limit
$\Sg\nearrow V_s$
\bea
\lft{\om^2 \int_{\Sg} G_\sg (\vr ,\vr ')\,v(\vr ')\,\vE(\vr ')\,d^3\vr'\rar}
\hs{1cm}\nn\\
&&-\oint_{\pa V_{s-}}\left[G_\sg (\vr ,\vr ')
(d\vS'\cdot\vnab')\vE(\vr') -\vE(\vr ')
(d\vS'\cdot\vnab')G_\sg (\vr ,\vr ')\right] \nn\\
&&
- \int_{V_s}\delta
(\vr-\vr ')\, \vE(\vr') \,d^3\vr' +\int_{V_{s-}} G_\sg (\vr ,\vr ')\vnab '
[\vnab '\cdot\vE(\vr')]\,d^3\vr'.\hs{0.8cm}
\lb{lab2}
\eea
Here, the l.h.s. of (\rf{lab2}) is well defined as a both-side limit.
Now, (\rf{entq}) can be rewritten on-shell as follows,
\bea
\lft{\vE(\vr) = \chi_{V_s}(\vr)\,\vE(\vr) +
\oint_{\pa V_{s-}}\left[G_\sg (\vr ,\vr ')\,(d\vS'\cdot\vnab')\vE(\vr ')-
\vE(\vr ')(d\vS'\cdot\vnab')\,G_\sg (\vr ,\vr ')
\right]}\hs{2cm}\nn\\
&&-\fr{1}{\eps_o}\oint_{\pa V_{s-}}\, [\vnab'G_\sg(\vr ,\vr ')]
v(\vr ')\,(\vE(\vr ')\cdot d{\bf S}')- \hs{3cm}\nn\\
&& - \oint_{\pa V_{s-}}G_\sg(\vr ,\vr ')
[\vnab '\cdot\vE(\vr ')]\,d\vS ',
\lb{elf1}
\eea
with $\chi_{V_s}(\vr)$ the characteristic function of $V_s$.

As for $\vB(\vr)$, equations (\rf{defb},\rf{inteq}) together with
(\rf{ME1}) imply that
\be
\vB (\vr) = \int_{V_{s+}}\fr{v(\vr ')}{\eps(\vr ')}\,
[\vnab ' G_\sg (\vr, \vr ')]\times
[\vnab '\times\vB(\vr ')]\,d^3\vr '.
\ee
By repeated use of the Gauss theorem one finds the desired on-shell result,
\bea
\lft{\vB(\vr) = \chi_{V_s}(\vr)\vB(\vr) +
\oint_{\pa V_{s-}}\fr{v(\vr')}{\eps(\vr')}\,G_\sg (\vr,\vr ')
d\vS '\times[\vnab\times\vB(\vr')]}\nn\\
&& +\oint_{\pa V_{s-}} \left\{
G_\sg(\vr, \vr ')\,(d\vS '\cdot\vnab ')\vB (\vr ')-
\vB(\vr ')\,
(d\vS '\cdot\vnab ') G_\sg (\vr,\vr ')\right\}.
\lb{ibelf}
\eea

We shall now compare outward (\rf{other}) and inward (\rf{elf1},
\rf{ibelf}) integral equations. For simplicity, we shall confine ourselves
to (\rf{elf1}).

\subsection{Relation of ``out" and ``in" formulations}
\lb{relation}
Here, we would like to derive (\rf{elf1}) from (\rf{other})
(or vice versa) by direct calculation of relation of $\vE(\vr)$ and
its normal derivative on different sides of $\Sg$ (see Appendix for
details).
For simplicity, we shall confine ourselves to such $\Sg$ for
which $(\pa_n\eps)_+ =(\pa_n\eps)_- =0$. Then the last surface integral
in (\rf{elf1}) vanishes.

As can be found in any textbook, equations (\rf{ME2})  or
(\rf{con}) imply the continuity of the normal component
$D_n(\vr)$ of $\vD(\vr)$ (i.\,e., a discontinuity of $E_n(\vr)$) or
the tangential component $E_t(\vr)$ of $\vE(\vr)$ on $\Sg$, respectively,
\be
E_n^+ - E_n^-=\fr{\eps_- -\eps_+}{\eps_+} E_n^-
=\fr{v(\vr)}{\eps_o}E_n^-, \hs{0.8cm}E_t^+=E_t^-.
\lb{uno}
\ee
On the other hand, the normal derivative $\pa_n \vE_t(\vr)$
of the tangential component of $\vE(\vr)$ changes discontinously across $\Sg$,
\be
\pa_n \vE_t^+ -\pa_n \vE_t^- =\left(\fr{1}{\eps_+}-\fr{1}{\eps_-}\right)
\vnab_t D_n =\fr{v}{\eps_o}\vnab_t E_n^-.
\lb{due}
\ee
Rather surprisingly, although the normal component $E_n(\vr)$ is
discontinuous,
\be
\pa_nE_n^+ =\pa_nE_n^-.
\lb{tre}
\ee

Now, by using (\rf{uno}) one can check directly the equivalence of
(\rf{other}) and (\rf{elf1}) for normal components. As for the tangential
components, one uses, assuming $\pa V_s$ to have no boundary, that
\bea
\lft{-\fr{1}{\eps_o}\oint_{\pa V_{s-}}\, [\vnab'G_\sg(\vr ,\vr ')]
v(\vr ')\,(\vE(\vr ')\cdot d{\bf S}') =}\hs{2.5cm}\nn\\
&&\fr{1}{\eps_o}\oint_{\pa V_{s-}}dS'\, G_\sg(\vr ,\vr ')
\vnab'[v(\vr ')E_n(\vr ')],
\eea
and
\be
\vnab_t [v(\vr)E_n(\vr)]=v(\vr)\vnab_t E_n(\vr).
\ee
The use of (\rf{due}) and (\rf{tre}) then gives the desired equivalence
of (\rf{other}) and (\rf{elf1}), as expected.

\section{Photonic KKR method}
\lb{kkr}
The KKR method is basically the method of rewriting {\em integral}
equations like (\rf{elf1},\rf{ibelf}) into {\em algebraic} ones
by using an appropriate basis (a basis of spherical harmonics in our case).
In the case of electrons it is known to lead to a very compact scheme if
the perturbing periodic potential
$v({\bf r})$ is {\em spherically symmetric} within inscribed
spheres, and zero  (constant) elsewhere \ct{KKR}.
The band structure of the problem is then determined by purely
{\em geometrical structure constants}, characteristic of the type of
lattice under considerations, and by {\em phase shifts}
(logarithmic derivatives, at the surface of the
inscribed sphere, of the $s,\,p,\,d\,\ldots$ radial
solutions of the corresponding radial equation),
characteristic of the
{\em scattering properties} of a given dielectric sphere.

In the case of photons the KKR method has been used  only
within the scalar approximation
to the Maxwell equations \ct{JR}. Recently, a version of the KKR
method has been given in the framework of {\em multiple-scattering
theory} \ct{XZ}. However, we have obtained a different  result \ct{AM1}.
Nevertheless, a derivation of the photonic KKR method in the
spirit of \ct{KKR} is still lacking.

Let us consider a dielectric lattice and look for  Bloch
wave-type solutions. In this case $G_\sg(\vr ,\vr ')$ possesses the standard
expansion in terms of the eigenfunctions of the homogeneous
boundary value problem,
\be
G_\sg (\vr ,\vr ')= -\,\fr{1}{\tau}\,\sum_n
\fr{\exp[i(\vK_n+\vk)\cdot(\vr -\vr ')]}
{(\vK_n +\vk)^2-\sg^2},
\lb{greenei}
\ee
where $\tau$ is the volume of the primitive cell $\Om$, $\vk$
is the Bloch momentum, and summation runs over all vectors
$\vK_n$ of a reciprocal lattice.

To derive the photonic KKR method we shall turn back to
either (\rf{elf1}) or (\rf{ibelf}).
There is one subtlety therein
with the characteristic function $\chi_{V_s}(\vr)$, as well as in (\rf{other}).
Depending on the order of the limits taken $\chi(\vr)$ may be
either one or zero for $\vr\in\pa V_{s}$. In order that $\chi(\vr)$
be zero  for $\vr\in\pa V_s$,
one takes as the first limit $\lim_{\vr\rar\pa V_s}$, and then
$\lim_{\Sg\rar V_{s-}}$ in (\rf{elf1}). In this case (\rf{elf1}) and
(\rf{ibelf})
give an integral equation for $\vE(\vr)$ and $\vB(\vr)$, respectively.
For a dielectric lattice of {\em spherical} atoms with radius $r_s$
it means that the expansion of $G_\sg(\vr,\vr')$ in spherical harmonics
$Y_{lm}(\theta,\phi)$ with $r_s>r>r'$ is used,
\bea
\lft{G_\sg(\vr ,\vr ') =\sum_{lm}\,\sum_{l'm'}\, [ A_{lm;l'm'}
j_l(\sg r')\,j_{l'}(\sg r)+}\hs{1cm} \nonumber\\
&& \mbox{}+ \sg \delta_{ll'}\delta_{mm'}n_l(\sg r)\,j_l(\sg r')]
\times \, Y_{lm}(\theta ',\phi ')\,Y^*_{l'm'}(\theta,\phi),\hs{2cm}
\lb{greengo}
\eea
where $j_l(x)$ is the {\em spherical Bessel function} defined by
\be
j_l(x) = \left(\fr{\pi}{2x}\right)J_{l+1/2}(x),
\ee
 $n_l (x) := (\pi/2x)^{1/2}J_{-l-1/2}(x)$ , and $\theta,\ \phi$ and
$\theta ',\ \phi '$ are polar angles of $\vr$ and $\vr '$, relative
to some fixed system of coordinates.
This expansion follows by expanding the exponentials in
(\rf{greenei}) for $G_\sg (\vr ,\vr ')$ \ct{KKR,J}.
The ``{\em structure constants}"
$A_{lm;l'm'}$, $A_{l'm';lm}=A_{lm;l'm'}{}^*$, which are functions of
$\sg$ and $\vk$, are
characteristic for the lattice under consideration. They are {\em exactly
the same} as for the case of electrons, i.e., as in the case of
the Schr\"{o}dinger equation.

For $r<r_s$ the true solution of our problem, for frequency $\om$,
can be expanded in vectorial spherical harmonics \ct{J},
\be
\vE(\vr) = \sum_{l=0}^\infty\,\sum_{m=-l}^{l}
\left[\fr{i}{\om\eps}\,C^{E}_{lm}\vnab\times\,[R^E_l(r)\,
\vX_{lm}(\theta,\phi)] + C^{M}_{lm}\,R^M_l(r)\,\vX_{lm}(\theta,\phi)\right],
\lb{mexp}
\ee
where the coefficients $C^{E}_{lm}$ and $C^{M}_{lm}$ specify
the amounts of {\em electric} $(l,m)$ and {\em magnetic} $(l,m)$
{\em multipole fields}.
$R^A_l(r)$, $A=E,\ M$, satisfies the corresponding radial part of equation
(\rf{ME*}) or (\rf{ME6*}) \ct{AK},
with the boundary conditions
\be
R^A_l(0)= {\rm finite},\hs{3cm}R^A_l(r_s)=1.
\lb{bc}
\ee
$R^M_l$ and $R^E_l$ differ in general if $\eps(r)\neq\, const$.
Now one takes scalar products of electric and magnetic multipoles
with both sides of (\rf{elf1}) and obtains a
matrix equation.
The condition of solvability, i.\,e., the vanishing of a corresponding
determinant, then determines dispersion relation and eventually photonic
bands.

On the other hand, the traditional variational derivation of the KKR
method from a variational principle \ct{KKR} uses the order of the
limits which makes $\chi_{V_s}(\vr)$ {\em zero} for $\vr\in\pa V_{s}$
(see (\rf{kkrf}) below).

\subsection{Variational principle for the Maxwell equations}
\lb{vari}
Let us formulate a general  variational
principle for the Maxwell equations which holds for an {\em arbitrary
shape} of the basic ``atom'' of a dielectric lattice.
Taking into account the off-shell integral equation (\rf{entq}) for $\vE(\vr)$
the off-shell photonic analogue of the scalar KKR functional is defined
to be
\bea
\lft{\Ld = \om^2\,\int_{V_s} d^3\vr\,v(\vr)\,\vE^*(\vr)\cdot \left\{\vE(\vr)
\mbox{} + \om^2\,
\int_{V_s} G_{\sg}(\vr ,\vr ')\,v(\vr')\,\vE (\vr ')d^3\vr ' +\right.}
\hs{4cm}\nonumber\\
&&\left. \mbox{} + \int_{V_{s+}} \vnab'
G_{\sg}(\vr ,\vr ') [\vnab'\cdot\vE (\vr ')] d^3\vr '
\right\}.
\lb{kkrf}
\eea
Here, the integral over $\vr$ is well defined and exists as
a {\em bothside} limit.
Provided $\eps(\vr)$ is {\em real} (no absorbtion), and
using well-known hermitian properties of Green functions,
\(\pa_{\vr}G_\sg (\vr ,\vr ')=-\pa_{\vr '}G_\sg (\vr ,\vr ')\) and
\(G^*_\sg(\vr ,\vr ') =G_\sg(\vr ',\vr)\), one can check that variations
of $\Ld$ with respect to $\vE(\vr)$ or $\vE^*(\vr)$ reproduce correctly the
equation for the electric field $\vE(\vr)$ (\rf{entq}) or its complex
conjugate $\vE^*(\vr)$ within $V_s$, respectively.
To deal properly with the singularities of $G$ we must use a
limiting procedure in evaluating $\Ld$. In analogy with \ct{KKR}  we set
\be
\Ld := \lim_{\eps\rightarrow 0}\, \Ld_\eps,
\ee
where schematically
\bea
\Ld_\eps := \om^2\,\int_{V_{s-2\eps}} d^3\vr\,v(\vr)\,\vE^*(\vr)
\cdot \left\{ \vE(\vr) +
\right.\hs{5cm} \nonumber\\
\left.\mbox{}+\om^2\,
\int_{V_{s-\eps}} G_{\sg}(\vr ,\vr ')\,v(\vr')\,\vE (\vr ')d^3\vr'
+ \int_{V_{s-\eps}} \vnab'
G_{\sg}(\vr ,\vr ')[\vnab'\cdot\vE(\vr ')]\,d^3\vr '\right.
\nn\\
\left.
\mbox{}+\fr{1}{\eps_o}\oint_{\pa V_{s-\eps}}
[\vnab'G_{\sg}(\vr,\vr ')]v(\vr ')(\vE(\vr')\cdot d\vS ')\right\}.
\lb{kkrfr}
\eea
Here, $V_{s-\eps}$ ($\pa V_{s-\eps}$) denotes the volume (boundary) of an
atom up to a shell of a width of $\eps$, one side of which is the boundary of
the atom and another is formed by the boundary shifted by $\eps$
in the direction of the inward normal.
Similarly for $V_{s-2\eps}$ and $\pa V_{s-2\eps}$.

Supposing that a complete set of interior solutions of
(\rf{ME*}) is known, it is more convenient
to find the on-shell KKR functional.  By using on-shell formulae
(\rf{lab2},\rf{elf1}) one finds the following on-shell form of the
variational KKR functional,
\bea
\lft{\Ld := \lim_{\eps\rightarrow 0}\left\{
\oint_{\pa V_{s-2\eps}} \left[[(d\vS\cdot\vnab)\vE^*(\vr)] -
\vE^*(\vr)(d\vS\cdot\vnab)
\right]-\right.}\hs{2cm}\nn\\
\lft{\left. \int_{V_{s-2\eps}} \vnab[\vnab\cdot\vE^*(\vr)]\,
d^3\vr\right\}\cdot \left\{\oint_{\pa V_{s -\eps}}\, \left[(d\vS '\cdot\vnab ')
\vE(\vr ')-\vE(\vr ')\,(d\vS '\cdot\vnab ')\right]G_\sg (\vr ,\vr ')
\right.}\hs{2cm}\nn\\
&&\left.\mbox{}-\fr{1}{\eps_o}\oint_{\pa V_{s-\eps}}\,[\vnab'G_\sg(\vr ,\vr
')]\,
v(\vr ')\,(\vE(\vr ')\cdot d{\bf S}')+\right.\hs{2.5cm} \nn\\
\lft{\left. - \oint_{\pa V_{s-\eps}}[G_\sg(\vr ,\vr ')\,
\vnab '\cdot\vE(\vr ')]\,d\vS' \right\}.}
\lb{final0}
\eea
Note that unless $\eps(\vr)$ is piecewise constant, i.\,e.,
the classical {\em muffin-tin potential}, it is {\em impossible}
(at least for our variational principle) to write
(\rf{kkrf}) on-shell in terms of surface integrals only.
Thus, when working with the variational KKR functional it is necessary
to confine ourselves to the case when $\eps(\vr)$ is
 a {\em real muffin-tin potential}. Then the dielectric
potential $v(r)=v$ can be moved in front of the integration sign,
the terms which contain $\vnab\cdot\vE(\vr)$ {\em vanish},
and (\rf{ME*}) simplifies within $V_s$ to
\be
(\vnab^2 +\rho^2)\,\vE(\vr) =0,
\lb{weq}
\ee
where $\rho=\om\sqrt{\eps}$.

In the case of spherically symmetric dielectric atoms
general formula (\rf{final0}) can be simplified further,
\bea
\lft{\Ld := \lim_{\eps\rightarrow 0}
\oint_{r=r_s-2\eps} dS\, \left[\pa_r \vE^*(\vr) -\vE^*(\vr)\pa_r
\right]\cdot}\hs{2cm}\nn\\
\lft{\cdot\left\{\oint_{r'=r_s -\eps}dS'\, \left[\pa_{r '}
\vE(\vr ')-\vE(\vr ')\,\pa_{r '}\right]G_\sg (\vr ,\vr ') -
\right.}\hs{2cm}\nn\\
&&\left.\mbox{}-\fr{1}{\eps_o}
\oint_{r'=r_s-\eps}\,[\vnab ' G_\sg(\vr ,\vr ')]\,
v(\vr ')\,(\vE(\vr ')\cdot d{\bf S}')\right\}.\hs{2.5cm}
\lb{final}
\eea
The final expression (\rf{final}) resembles
the scalar case \ct{KKR}, the only difference being the last term.

{}From now on, a further treatment along the lines in \ct{KKR}
is straightforward, albeit more involved.
The true solution of our problem, for frequency $\om$
and for $r<r_s$, is approximated by finite sums
of the form (\rf{mexp}) with $R^A_l(r)$
being now $R^A_l(r) =a_l\, j_l(\rho r)$, $a_l$ being
a {\em normalization constant}
(see for example \ct{J}). Then, as in the previous case,
bands are obtained from the condition of solvability which requires that the
determinant of $\Ld$ be zero.
%

\section{Conclusions}
We have presented and discussed  the relationship between
the outward and inward integral equations for electromagnetic
waves and we have outlined a derivation
of the KKR method. For a correct implementation of the photonic KKR
method in the spirit of \ct{KKR},
inward on-shell integral equations for electromagnetic fields
are necessary.
The photonic KKR method then follows from them either directly
or through a variational principle.
The ``direct"  photonic KKR method has a wider region of applications :
complex and non-constant $\eps(\vr)$ within atoms.
Rather surprisingly, we have found that the variational KKR method
has a rather limited range of applications. Unless $\eps(\vr)$ is piecewise
constant the variational KKR method  {\em cannot} be written
in terms of surface integrals. Also, for complex $\eps(\vr)$,
the variational KKR method cannot be used. However, being variational
it is expected to converge more rapidly within its range of application.
Both of the photonic KKR methods presented make full use of the
same structure constants as the electronic KKR method.
We hope soon to report on numerical results \ct{AK}.

In our formulation of the photonic KKR method singularities due to
discontinuities of permeabilities are {\em safely} treated.
Our results also imply that the photonic KKR method as proposed in
\ct{LWA} can be only used if, at the atom boundary
permeabilities are {\em continuous} and in addition side limits of their
derivatives are identically zero. One's first thought may be to define the
atom boundary to comprise discontinuities of permeabilities such that
the photonic KKR method of \ct{LWA} can be used. However, afterwords,
singularities of delta function are moved into coefficients of
corresponding radial equations.

Multiple-scattering theory for photons, which uses essentially outward
integral equations, is presented in \ct{AM1}.

\section{Acknowledgments}
I should like to thank Prof. H. Kunz for suggesting this problem
and for valuable discussions
during my postdoctoral stay at IPT EPFL Lausanne. Financial support
of the Swiss National Foundation during the early stages of this work,
and partial support by the CAS grant No. 11086, are also gratefully
acknowledged.

\appendix
\section{Appendix : Derivatives of $\vE(\vr)$ on different
sides of a normal discontinuity}
\lb{disco}
Let $\Sg$ be some surface discontinuity of $\eps(\vr)$ and $\mu(\vr)=1$
in $\Om$.
We shall assume that near $\Sg$ only the normal derivative of $\eps(\vr)$ may
be non-zero. In what follows, such a discontinuity will be called {\em normal}.
In order to determine the discontinuity of $\pa_n \vE_t(\vr)$
we shall look carefully at the tangential components of (\rf{ME*}),
\be
\vnab^2\vE_t(\vr)-\pa_n\eps(\vr)^{-1}\vnab_t D_n(\vr) +
\om^2\eps(\vr)\vE_t(\vr)=0,
\lb{pufi}
\ee
where we have used \(\vnab\cdot\vE(\vr)=\vD(\vr)\cdot\vnab\eps(\vr)\).
The most singular terms are given by the normal derivatives of $\eps(\vr)$.
Obviously,
\be
\left.\vnab\fr{1}{\eps(\vr)}\right|_\Sg =
\left[\fr{1}{\eps_+}-\fr{1}{\eps_-}\right]\delta(\vr-\Sg)\,\vn.
\ee
The requirement of cancellation of singular terms gives
$\pa_n\vE_t(\vr)=\eps^{-1}(\vr)\vnab_t D_n(\vr)$, i.\,e.,
\be
\left.(\pa_n \vE_t^+ -\pa_n \vE_t^-)\right|_\Sg =\left(\fr{1}{\eps_+}-
\fr{1}{\eps_-}\right) \vnab_t D_n =\fr{v}{\eps_o}\vnab_t E_n^-.
\ee
Since $\vnab_t D_n(\vr)$ is continuous (see below) one also finds
$\pa_n\vE_t^-=\vnab_t E_n^-$.

On the other hand, to determine the relation between normal derivatives of
the normal component of $\vE(\vr)$  on different sides of
the discontinuity
$\vE(\vr)$ is replaced by $(\vD(\vr)/\eps(\vr))$ in (\rf{ME*}). After
some manipulations one finds (for the normal component)
\be
\vnab^2 D_n+\eps(\vr)\left(\pa_n\fr{1}{\eps(\vr)}\pa_n\right)D_n+
\om^2\eps(\vr) D_n=0.
\lb{mefi}
\ee
Note that terms proportional $\vnab^2(1/\eps(\vr))$ have cancelled.
Now, the requirement of cancellation of remaining singular terms
in (\rf{mefi}) gives conditions on $\pa_nD_n(\vr)$ and $\vnab_t D_n(\vr)$.
$\vnab_t D_n(\vr)$ changes continuously across $\Sg$ while $\pa_nD_n(\vr)$
has to be discontinuous.
The discontinuity is such that $\pa_nD_n(\vr)/\eps(\vr)$ change
continuously, i.\,e.,
\be
\fr{\pa_nD_n^+}{\eps_+}=\fr{\pa_nD_n^-}{\eps_-}\cdot
\ee
This in turn implies that
\be
\left.(\pa_nE_n^+ -\pa_nE_n^-)\right|_\Sg =-
\left(\fr{1}{\eps_+^2}(\pa_n\eps)_+ -\fr{1}{\eps_-^2}(\pa_n\eps)_-\right)D_n.
\ee
Thus,  although the normal component $E_n(\vr)$ of $\vE(\vr)$ is discontinuous,
the side limits of $\pa_nE_n(\vr)$ {\em coincide} provided the side limits of
the derivatives of $\eps(\vr)$ at $\Sg$ are zero.

Note that the above relations are a general consequence of
the (non-stationary) Maxwell equations in a dielectric.
\np

\end{document}